\pdfoutput=1
\documentclass[prl,aps,twocolumn]{revtex4-2}
\usepackage[colorlinks=true,citecolor=blue,urlcolor=blue,linkcolor=blue,pdfstartview=FitH,bookmarksopen]{hyperref}
\usepackage{amsmath,amssymb}
\usepackage{bm}
\usepackage{graphicx} 
\usepackage[colorlinks=true,citecolor=blue,urlcolor=blue,linkcolor=blue,pdfstartview=FitH,bookmarksopen]{hyperref}
\usepackage{xcolor}

\usepackage{comment}

\newcommand\+\dagger
\renewcommand\d\partial

\newcommand{\diff}{\mathrm{d}}

\begin{document}

\title{Theory of Spinful Relativistic Superfluids}
\author{Dam Thanh Son}
\affiliation{Leinweber Institute for Theoretical Physics, University of Chicago, Chicago, Illinois 60637, USA}

\date{July 2026}

\begin{abstract}
We construct an effective field theory description of relativistic
superfluids with nonzero angular momentum density.  At first order in
the derivative expansion, the effective action contains a term with a
quantized coefficient, which encodes the Berry phase for the angular
momentum of the superfluid condensate.  From this single term we
derive a number of physical effects, including the relativistic
Mermin-Ho relation, an anomalous Ettingshausen effect (exchangeable
for an anomalous Hall effect), and an anomalous Hall viscosity.
\end{abstract}

\maketitle

\emph{Introduction.}---Since its discovery almost 90 years ago in
liquid helium,
the superfluid state~\cite{Khalatnikov:1965} has appeared in diverse
physical environments, ranging, in terms of the energy scale involved,
from the hypothetical ultralight axion dark matter~\cite{Hui:2016ltb}
to trapped ultracold atoms~\cite{Davis:1995pg,Anderson:1995gf} to the
core of neutron stars (as nuclear matter~\cite{Hoffberg:1970vqj} or
quark matter superfluids~\cite{Alford:1998mk}).  The gauged
counterpart of the superfluids---superconductors---also have many
realizations in solid states, nuclear and quark matter.  When the
chemical potential is considerably higher than the rest mass of the
particles that make up the superfluid, as in quark matter and
high-density nuclear matter, relativistic effects become considerable,
and to describe such superfluids a relativistic version of the
Landau-Khalatnikov hydrodynamic theory~\cite{Khalatnikov:1965} was
developed in the 1980s~\cite{Lebedev:1982,Carter:1992gmy}.  At zero
temperature, the structure of the low-energy effective theory
simplifies drastically: the only low-energy degree of freedom is the
U(1) phase of the symmetry-breaking condensate, and the effective
action, to leading order of the derivative expansion, is completely
fixed by the zero-temperature equation of state~\cite{Son:2002zn}.

Previous works on the theory of relativistic superfluids assume a
scalar order parameter, i.e., one considers only $s$-wave superfluids,
similar to superfluid $^4$He.  However, it has been suggested that
nuclear and quark matter may exist in a superfluid or superconducting
state where the Cooper pair carries nonzero angular momentum, as in
superfluid $^3$He~\cite{VollhardtWoelfle-book}.  For example, while at
low density neutrons pair in $s$-wave, at high density the preferred
pairing channel is $^3P_2$~\cite{Hoffberg:1970vqj}.  In quark matter,
a single quark flavor should also pair in $p$-wave.  To describe the
low-energy dynamics of these systems, one needs to take into account
the extra Nambu-Goldstone bosons associated with the direction of the
angular momentum of the Cooper pairs.

In this Letter we develop the effective field theory of spinful
relativistic superfluids, by which we mean those whose ground state
carries nonzero density of angular momentum.  One feature that makes
these superfluids interesting is that the effective field theory
describing them contains a topological term with a quantized
coefficient responsible for a number of parity-odd transport
phenomena, all with rigidly fixed kinetic coefficients.  Nonzero
angular momentum density occurs in the $A$-phase of superfluid $^3$He
and, e.g., in region I in Mermin's classification of minima of the
Ginzburg-Landau functional for $d$-wave paring~\cite{Mermin:1974zz},
or one of the magnetized phases considered in
Ref.~\cite{Mizushima:2021qrz}. A neutron superfluid with such a paring
pattern would be ferromagnetic, and if realized, may provide a source
for the the large magnetic fields of magnetars.  It is not guaranteed,
however, that nuclear interactions favor such a ferromagnetic state at
any density.
In the standard treatment, the ground state of the $^3P_2$ neutron
superfluid is not ferromagnetic~\cite{Hoffberg:1970vqj}, but a
ferromagnetic phase may become stable in a range of magnetic
field~\cite{Mizushima:2021qrz}.  In this paper will focus on deriving
the low-energy physics of a spinful superfluid, leaving aside the
question of whether and where such a state occurs in nature.

\emph{Dual-variable action for a $s$-wave superfluids}.---We first
review the theory of $s$-wave superfluids.  In the standard theory,
the effective Lagrangian is written in terms of a scalar field---the
U(1) phase of the condensate, but to extend to the spinful case
another description in terms of the dual two-form gauge field is more
convenient. In this dual description, the fundamental field is a
two-form gauge field $b_{\mu\nu}=-b_{\nu\mu}$; one requires invariance
under the gauge transformation
\begin{equation}
  b_{\mu\nu} \to b_{\mu\nu} + \d_\mu\alpha_\nu - \d_\nu \alpha_\mu .
\end{equation}
The gauge-invariant three-form field strength,
\begin{equation}
  h_{\mu\nu\lambda} = \d_\mu b_{\nu\lambda} + \d_\nu b_{\lambda\mu}
    + \d_\lambda b_{\mu\nu}, 
\end{equation}
is postulated to be proportional to the U(1) number
current~\footnote{We use the $(-+++)$ metric convention.  Here
$\epsilon^{\mu\nu\lambda\rho}$ is the Levi-Civita tensor,
$\epsilon^{0123}=1$. We will not concern ourselves with the quantized
nature of the U(1) charge.}
\begin{equation}
  h_{\mu\nu\lambda} = \epsilon_{\mu\nu\lambda\rho} j^\rho, \quad
  j^\mu = \frac16 \epsilon^{\mu\nu\lambda\rho} h_{\nu\lambda\rho} .
\end{equation}
From the current $j^\mu$ one defines the number density and
four-velocity,
\begin{equation}
  j^\mu = nu^\mu, \quad u^\mu u_\mu = -1,
\end{equation}
hence
\begin{equation}
  n = \left( \frac16 h_{\mu\nu\lambda} h^{\mu\nu\lambda}\right)^{1/2}, \quad
  u^\mu = \frac 1{6n} \epsilon^{\mu\nu\lambda\rho} h_{\nu\lambda\rho} .
\end{equation}
In the dual description, the continuity equation $\d_\mu (nu^\mu)=0$
automatically follows from the definition of the current.

The action for a superfluid is
\begin{equation}\label{eq:S-swave}
  S = - \int\! \mathrm d^4x\, \epsilon(n),
\end{equation}
where $\epsilon(n)$ is the energy density as a function of the number
density.  Varying the action~(\ref{eq:S-swave}) one obtains
\begin{equation}\label{eq:irrot0}
  \d_\lambda (\mu u_\rho) - \d_\rho(\mu u_\lambda) = 0,
\end{equation}
where $\mu=\epsilon'(n)$ is the chemical potential.  This is the
irrotationality condition for relativistic superfluid flows: in the
absence of vortices, the ``specific four-momentum'' (also called the
``enthalpy current'') $\mu u_\mu$ is a closed one-form.
Equation~(\ref{eq:irrot0}) and the continuity equation form the full
set of equations of zero-temperature superfluid hydrodynamics.

Coupling to the external U(1) gauge field can be done by the inclusion
of the term $A_\mu j^\mu=\frac12 \epsilon^{\mu\nu\lambda\rho}A_\mu
\d_\nu b_{\lambda\rho}$ in the Lagrangian density.

\emph{A closed two-form for a spinful fluid.}---We now consider a
spinful superfluid.  We assume that at each spacetime point $x$ of
which there is a unit timelike vector $u^a$, identified with the
direction of the U(1) particle number current, and a unit spacelike
vector $S^a$, which is the direction of the spin of the condensate
(e.g., the Cooper pair)~\footnote{We ignore the possible nodal
fermions.  When these fermions appear at specific points on the Fermi
surface, their density of state scales like the square of energy and
thus the effects of these fermions show up only at the two-derivative
order, i.e., one order beyond the precision of this Letter.},
\begin{equation}\label{eq:uS-ortho}
   u^a u_a = -1, \quad
  S^a S_a = 1, \quad
  u^a S_a = 0,
\end{equation}
For clarity of notation we use the indices $a$, $b$, etc., for the
tangent vectors $u^a$ and $S^a$; in flat space these indices are
identical to the Lorentz indices $\mu$, $\nu$, etc.

A central object of our construction is a two-form, $J= J_{\mu\nu}
\diff x^\mu \wedge \diff x^\nu$, where
\begin{equation}\label{eq:J}
  J_{\mu\nu} =
  \epsilon_{abcd}
   u^a S^b (\d_\mu S^c \d_\nu S^d - \d_\mu u^c \d_\nu u^d).
\end{equation}
A few immediate remarks can be made.  First, this two-form is
invariant under spacetime-dependent Lorentz boosts,
\begin{subequations}\label{eq:local-Lorentz}
\begin{align}
  u^a &\to \cosh\beta(x) u^a + \sinh\beta(x) S^a ,\\
  S^a &\to \cosh\beta(x) S^a + \sinh\beta(x) u^a .
\end{align}
\end{subequations}
In other words, $J_{\mu\nu}$ depends only on the two-dimensional plane
spanned by $(u^a, S^a)$, but not on the individual basis vectors $u^a$
and $S^a$ on that plane.  This fact can be made explicit by expressing
$J_{\mu\nu}$ through the antisymmetric tensor $L^{ab}=u^a S^b- S^a
u^b$ characterizing the plane [which is invariant
under~(\ref{eq:local-Lorentz})],
\begin{equation}
  J_{\mu\nu} = \frac12 \epsilon_{abcd} L^{ab} \d_\mu {L^c}_e \d_\nu L^{ed} .
\end{equation}
Second, when we fix the fluid four-velocity to be along the time
direction, $u^a=(1,\mathbf 0)$, the two-form $J$ is the Skyrmion
current associated with the spin configuration $S^i(x)$,
\begin{equation}
  J_{\mu\nu} = - \epsilon_{ijk} S^i\d_\mu S^j \d_\nu S^k .
\end{equation}
And finally, if one freezes the spin vector to the $z$-direction,
$S^\mu=(0,0,0,1)$, then $\star J$ is the Euler
current~\cite{Golkar:2014paa} from the point of view of the (2+1)D
space $(t,x,y)$.

By direct calculation, one can show
that the two-form $J$ is
closed: $\diff J=0$, or, in components,
\begin{equation}
  \d_\mu J_{\nu\lambda} + \d_\nu J_{\lambda\mu} + \d_\lambda J_{\mu\nu} = 0 .
\end{equation}
Equivalently, $\star J$ is a conserved antisymmetric tensor
current,
\begin{equation}
   \d_\nu (\star J)^{\mu\nu} \equiv \frac12 \epsilon^{\mu\nu\lambda\rho}\d_\nu J_{\lambda\rho} = 0.
\end{equation}

The two-form $J$ has a geometric interpretation.  The manifold of all
pairs of perpendicular timelike and spacelike unit vectors $u^a$ and
$S^a$ in (3+1)D Minkowski space is the coset $SO(3,1)/SO(2)\sim
SO(1,1)\times S^2$.  $J$ is the pullback of the volume two-form on the
$S^2$ to the physical space.  Since the volume form is a top form on
$S^2$, $J$ is guaranteed to be closed.

\emph{Action for spinful superfluids.}---Since the current
$J_{\mu\nu}$ is closed, one can add to the action a new term
\begin{equation}
  S = \int\!\mathrm d^4x\, \left[ -\epsilon(n)
  + \frac{s}4 \epsilon^{\mu\nu\lambda\rho} b_{\mu\nu} J_{\lambda\rho} \right].
\end{equation}
The Lagrangian density in the second term is not gauge invariant, but
thanks to $J$ being closed, its change under gauge transformation is a
total derivative.  Note that gauge invariance requires that $s$ is a
constant and cannot be a nontrivial function of the density $n$.  Thus
the value of $s$ is a topological property of the superfluid phase.

The natural power counting scheme is the one where $h_{\mu\nu\lambda}$
and $S_\mu$ are considered $O(p^0)$ in the derivative expansion
[which means $b_{\mu\nu}\sim O(p^{-1})$].  The term proportional to
$s$ is suppressed by one derivative compared to the leading term.
There is only one more $O(p)$ term consistent with parity and time
reversal, the effects of which will be consider at the end of this
paper.

To understand the physics behind the new term, let us consider the
special case when the fluid is at rest, $u^a=(1,\mathbf 0)$, so $S^a$
is a purely spatial vector, $S^a=(0, \mathbf S)$.  Parametrizing
$\mathbf S$ using the polar coordinates $\theta$ and $\phi$, we have
\begin{equation}
  J_{\mu\nu} = -\epsilon_{abc}S^a \d_\mu S^b \d_\nu S^c 
  = - \sin\theta \d_\mu\theta \d_\nu\phi ,
\end{equation}
and, by integration by part, the new term in the action is brought to the form
\begin{equation}
  S = -\int\!\mathrm d^4x\, s n(\mathbf x) (1-\cos\theta) \d_t \phi .   
\end{equation}
This action can be interpreted as the Berry phase of the spin degrees
of freedom, with $sn(\mathbf x)$ being the density of spin.  Thus $s$
is the spin per particle.  For fermionic superfluids, $s$ is half of
the angular momentum of the Cooper pair.

\emph{Equations of motion.}---To obtain the equations of motion, one
needs to vary the action.  For that, it is useful to have the formula
for the variation of the two-form $J_{\mu\nu}$ under variations
$u^a \to u^a+\delta u^a$ and $S^a \to S^a+\delta S^a$ that preserves
the constraints~(\ref{eq:uS-ortho}):
\begin{equation}
  \delta J_{\mu\nu} = \d_\mu K_\nu - \d_\nu K_\mu,
\end{equation}
where
\begin{equation}
  K_\mu = \epsilon_{abcd} u^a S^b (\d_\mu u^c \delta u^d
    - \d_\mu S^c \delta S^d).
\end{equation}

Varying the action with respect to $S^a$ we find the equation of
motion for the spin,
\begin{equation}
  \epsilon_{abcd} u^b S^c (u\cdot \d)S^d = 0.
\end{equation}
This means that the vector $(u\cdot\d)S^a$ lies in the plane spanned
by $u^a$ and $S^a$.  Since $S^a$ has unit length, $(u\cdot \d)S^a\sim
u^a$.  The coefficient in this term can be determined from the
condition $S\cdot u=0$.  In this way the equation of motion for the
spin is found to be
\begin{equation}\label{eq:S-dot}
  (u\cdot\d) S^a + u^a S_b (u\cdot \d)u^b =0,
\end{equation}
implying that spin is Fermi-Walker transported along the flow line.
To see a more nontrivial evolution of the spin, one needs to include
into the Lagrangian second-derivative terms like $\d_\mu S^a \d_\mu
S^a$.

Varying the action with respect to $b_{\mu\nu}$ we get
\begin{equation}\label{eq:irrot}
  \d_\lambda (\mu u_\rho + C_\rho) - \d_\rho (\mu u_\lambda + C_\lambda)
  + s J_{\lambda\rho} = 0,
\end{equation}
where
\begin{equation}
  C_\mu = - s\epsilon_{\mu\alpha\beta\gamma} u^\alpha S^\beta
          (u\cdot\d)u^\gamma .
\end{equation}
We see here two modifications to the condition of irrotationality of
the superfluid flow: (i) there is a correction $C_\mu$ to the specific
four-momentum, and (ii) the vorticity is not equal to zero but to the
2-form $J$.  The correction $C_\mu$ is a relativistic effect: by
restoring the speed of light $c$ in Eq.~(\ref{eq:irrot}) one can see
that $C_\rho$ comes with a factor of $c^{-2}$ where $c$ is the speed
of light.  In the nonrelativistic limit the spatial components of
Eq.~(\ref{eq:irrot}) give the Mermin-Ho relation, first found in the
$A$-phase superfluid $^3$He~\cite{Mermin:1976zz}, and according to
which Skyrmion lines carry superfluid vorticity.

Together with the continuity equation $\d_\mu (nu^\mu)=0$,
Eqs.~(\ref{eq:S-dot}) and (\ref{eq:irrot}) are the equations of motion
of a spinful superfluid.

\emph{Stress-energy tensor}.---From the equations of motion, one can show
that the following stress-energy tensor
\begin{equation}
  {T^\mu}_\nu = (\epsilon+P) u^\mu u_\nu + P\delta^\mu_\nu + n u^\mu C_\nu,
\end{equation}
is conserved, $\d_\mu {T^\mu}_\nu = 0$.  This stress tensor, with both
indices raised, is not symmetric: $T^{\mu\nu}\neq T^{\nu\mu}$.  We search
for a spin current $\Sigma^{\mu\nu\lambda}$ which satisfies the equation
\begin{equation}
  \d_\mu \Sigma^{\mu\alpha\beta} = - T^{\alpha\beta} + T^{\beta\alpha} .
\end{equation}
This following current can be shown to satisfy the above condition:
\begin{equation}
  \Sigma^{\mu\alpha\beta} = -s n u^\mu \epsilon^{\alpha\beta\gamma\delta} u_\gamma S_\delta .
\end{equation}
The current points along the direction of fluid motion.  In the fluid
rest frame, the spin density is $\Sigma^{0ij}=sn\epsilon^{ijk}S_k$.

Using the spin current, one can construct the symmetric
(Belinfante-Rosenfeld) stress energy tensor,
\begin{equation}
  T^{\mu\nu}_\text{sym} = T^{\mu\nu} + \frac12\d_\lambda
  (\Sigma^{\lambda\mu\nu} - \Sigma^{\mu\lambda\nu} - \Sigma^{\nu\lambda\mu}).
\end{equation}
Then one finds $T^{\mu\nu}_\text{sym}=(\epsilon+P)u^\mu u^\nu +
Pg^{\mu\nu} + \delta T^{\mu\nu}$, where the one-derivative correction
$\delta T^{\mu\nu}$ can be decomposed as
\begin{equation}\label{eq:Tepsilonqtau}
  \delta T^{\mu\nu} = \delta \epsilon u^\mu u^\nu + u^\mu q^\nu + u^\nu q^\mu
   + \tau^{\mu\nu},
\end{equation}
where $u_\mu q^\mu= u_\mu \tau^{\mu\nu}=0$.  Here we give the
expressions for the quantities that appear in
Eq.~(\ref{eq:Tepsilonqtau}) in the local fluid rest frame, where
$u^\mu=(1,\mathbf 0)$ at the given point,
\begin{subequations}\label{eq:epsilonqtau}
\begin{align}
  \delta \epsilon &=  - sn \epsilon^{ijk}S_i \d_j v_k, \label{eq:T00}\\
  q^i &= \frac 12 \epsilon^{ijk}\d_j (snS_k) +sn \epsilon^{ijk}
  \d_t v_j S_k, \label{eq:T0i}\\
  \tau^{ij} &=  \frac {sn}2 (\epsilon^{ikl}\d_k v^j +
  \epsilon^{jkl}\d_k v^i) S_l . \label{eq:Tij}
\end{align}
\end{subequations}

These formulas can be given physical interpretation.  The correction
to the energy density $\delta\epsilon$ can be written as
$\delta\epsilon=-2\bm{\sigma}\cdot\bm{\omega}$ where
$\bm{\sigma}=sn\mathbf S$ is the spin density and $\bm{\omega}
=\frac12\bm{\nabla}\times\mathbf v$ the vorticity of the flow.  This
formula has the form of the Mashhoon effect~\cite{Mashhoon:1988zz}
with a gravitational gyromagnetic factor of 2.  There are two terms in
Eq.~(\ref{eq:T0i}) for the correction to the momentum density or
energy current $\mathbf q$.  The first contribution,
$\frac12\bm\nabla\times\bm\sigma$, is the ``magnetization'' momentum
density, analogous to the magnetization current.  The second
contribution, by using leading-order equation of motion [in external
U(1) gauge field] $\mu \d_t v_i = E_i-\d_i \mu$, can be identified
with the anomalous Ettingshausen effect, i.e., an energy flux in the
direction perpendicular to the electric field and the spin density,
\begin{equation}
  \mathbf q_\text{AE}^{\phantom{1}} = \Pi_\text{AE} (\mathbf E-\bm\nabla\mu) \times \bm\sigma ,
\end{equation}
where the anomalous Ettingshausen coefficient $\Pi_\text{AE}=1/\mu$.
Note that we have been using the Eckart frame where $j^\mu=nu^\mu$
exactly.  In the Landau-Lifshitz frame, instead of the anomalous
Ettingshausen effect one has an anomalous Hall contribution to the
current, $\mathbf j=\cdots+ \sigma_\text{AH}(\mathbf
E-\bm\nabla\mu)\times\bm\sigma$, with the anomalous Hall coefficient
$\sigma_\text{AH}=-1/\mu^2$.  In a general frame, both $\Pi_\text{AE}$
and $\sigma_\text{AH}$ are nonzero, but the linear combination
$\Pi_\text{AE}-\mu\sigma_\text{AH}$ is frame-independent and is equal
to $1/\mu$.

Finally, the first-order corrections to the stress,
Eq.~(\ref{eq:Tij}), can be rewritten as
\begin{equation}
  \tau_{ij} = \frac14 (\epsilon_{ikl} V_{jk} + \epsilon_{jkl}V_{ik})\sigma_l + \frac12
  (\omega_i \sigma_j + \omega_j
  \sigma_i - 2\delta_{ij}\bm{\omega}\cdot \bm\sigma),
\end{equation}
where $V_{ij}\equiv \d_i v_j + \d_j v_i$.  The first term can be
called anomalous odd (or Hall) viscosity~\cite{Avron:1995fg}, while
the second term is an anisotropic contribution to the stress tensor
that comes from the interplay between rotation and spin.  Note that
$\delta\epsilon - \tau_{ii}=0$, i.e., the first-order correction to
the stress-energy tensor is traceless.

\emph{Effective field theory in curved space.}---One can consistently
couple the effective field theory to a metric.  By direct calculation we
can show that the two-form with components
\begin{equation}
  J_{\mu\nu} = \epsilon_{\alpha\beta\gamma\delta} u^\alpha S^\beta \Bigl( \nabla_\mu S^\gamma\nabla_\nu S^\delta - \nabla_\mu u^\gamma\nabla_\nu u^\delta  + \frac12 {R_{\mu\nu}}^{\gamma\delta}\Bigr) 
\end{equation}
with ${R_{\mu\nu}}^{\gamma\delta}$ being the Riemann tensor
of spacetime, is closed.  One can again add a term $\frac14 s
\epsilon^{\mu\nu\lambda\rho} b_{\mu\nu} J_{\lambda\rho}$ to the action
and obtain an action that is both gauge and general-coordinate
invariant.  The response of the system to metric perturbations can be
computed from this action.  In particular, one finds transverse
response corresponding to the odd viscosity.

\emph{Magnetic moment.}---Beside the topological term, at the
one-derivative level, there is one more term one may add to the
effective Lagrangian which is invariant under both parity and time
reversal:
\begin{equation}
  \lambda(n) \epsilon^{\mu\nu\lambda\rho} u_\mu \d_\nu u_\lambda S_\rho,
\end{equation}
where $\lambda(n)$ is any function of density.  
This term can be absorbed into the leading order term in the Lagrangian $-\epsilon(n)$ by using a freedom in redefining $b_{\mu\nu}$, i.e., by changing
\begin{equation}
  b_{\mu\nu} \to b_{\mu\nu} - M(n) (u_\mu S_\nu - u_\nu S_\mu),
\end{equation}
with $M(n)=\lambda/\mu$. The effect of this change of
variable is a nonminimal coupling to the external U(1) gauge field,
\begin{equation}
  -\frac12 M(n) \epsilon^{\mu\nu\lambda\rho} F_{\mu\nu} u_\lambda S_\rho,
\end{equation}
so $M(n)$ can be identified with the magnetic moment density.  The new
term does not modify the equations of motion in the absence of the
external U(1) gauge field, but modifies the U(1) current by adding to
it a divergenceless magnetization current,
\begin{equation}
  j^\mu = nu^\mu -\epsilon^{\mu\nu\lambda\rho} \d_\nu [M(n) u_\lambda S_\rho].
\end{equation}
In the absence of external fields, all other formulas derived
previously remain unchanged.

\emph{Conclusion.}---We have developed an effective field theory
describing that spinful relativistic superfluids.  The effective field
theory involves one term with quantized coefficient, from which
follows a number of physical phenomena.

In this paper, we have limited ourselves to one-derivative corrections
to the effective Lagrangian.  It should be possible to list all
two-derivative terms consistent with symmetries.  As mentioned above,
terms like $\d_\mu S^\alpha \d^\mu S_\alpha$ would give the spin waves
a quadratic dispersion relation of a type-2 Nambu-Goldstone boson.

It would be interesting to investigate the properties of defects in
the theory.  An curious feature is that the Skyrmion line carries
vorticity.  Its interaction with the Nambu-Goldstone bosons may be
nontrivial.  Finite-temperature (two-component) hydrodynamic theory of
a relativistic spinful superfluid is still to be constructed.  We
expect that theory to share many common features with (normal)
relativistic hydrodynamic with spin, which is being intensively
studied~\cite{Li:2020eon,Hongo:2021ona,Huang:2024ffg}.

In complete analogy with the theory constructed here, one can write
down a theory of a relativistic solid with spin.  In effective field
theory, a solid is described as a map from the external coordinates
$x^\mu$ to the internal spatial coordinate $X^a$,
$a=1,2,3$~\cite{Soper-book}.  The U(1) current is $j^\mu\sim
\epsilon^{\mu\nu\lambda\rho} \epsilon_{abc} \d_\nu X^a \d_\lambda X^b
\d_\rho X^c$; the following topological term can thus be added to the
Lagrangian
\begin{equation}
  \epsilon^{\mu\nu\lambda\rho} \epsilon_{abc} X^a\d_\mu X^b \d_\nu X^c
  J_{\lambda\rho} .
\end{equation}

Finally, we note that, in the same way as parity-odd (2+1)D superfluid
hydrodynamics is useful in the large-charge sector of certain 3D
conformal field theories (CFTs)~\cite{Cuomo:2021qws}, the effective
field theory described in this paper may find applications in the
high-charge sector of (3+1)D
CFTs~\cite{Hellerman:2015nra,Monin:2016jmo}.

\acknowledgments

The author thanks Clay C\'ordova, Keisuke Harigaya, Marvin Qi, Mikhail
Stephanov, and Ho-Ung Yee for discussions.  This is supported, in
part, by the U.S.\ DOE grant No.\ DE-FG02-13ER41958 and by the Simons
Collaboration on Ultra-Quantum Matter, which is a grant from the
Simons Foundation (No.\ 651442, DTS).

\bibliography{ferrofluids}

\end{document}